\documentclass[twoside]{article}

\usepackage[sc]{mathpazo} % Use the Palatino font
\usepackage{microtype} % Slightly tweak font spacing for aesthetics

\usepackage{mhchem}
\usepackage[hmarginratio=1:1, top=25 mm, left=25mm, columnsep=20pt]{geometry} % Document margins
\usepackage{multicol} % Used for the two-column layout of the document
\usepackage{float}
\usepackage[small,labelfont=bf,up,textfont=it,up]{caption} % Custom captions under/above floats in tables or figures
\usepackage{booktabs} % Horizontal rules in tables
\usepackage{float} % Required for tables and figures in the multi-column environment - they need to be placed in specific locations with the [H] (e.g. \begin{table}[H])
\usepackage[section]{placeins}

\usepackage{url}

\usepackage{xy}
\usepackage{graphicx}
\graphicspath{{rys/}{wykresy/}{foto/}}
\DeclareGraphicsExtensions{.png, .PNG, .jpg, .JPG, .jpeg, .JPEG, .gif, .GIF, .tiff, .TIFF, .TIF, .tif, .PDF, .pdf}

\usepackage{paralist} % Used for the compactitem environment which makes bullet points with less space between them

\usepackage{abstract} % Allows abstract customization
\usepackage{titlesec} % Allows customization of titles
\titleformat{\title}         % Customise the \section command
  {\huge} % Make the \section headers large (\Large),
  {}{0em}                      % Can be used to give a prefix to all sections, like 'Section ...'
  {}                           % Can be used to insert code before the heading
\titleformat{\section}[block]{\Large\scshape\raggedright}{\thesection.}{0.5em}{}[\titlerule] % Change the look of the section titles
\titleformat{\subsection}[block]{\large\scshape}{\thesubsection.}{0.5em}{} % Change the look of the section titles

\usepackage{fancyhdr} % Headers and footers
\usepackage{xfrac}

\begin{document}

%\twocolumn[
\title{\vspace{-15mm}\fontsize{24pt}{24pt}\selectfont\textsc{Fast minute magnetic field coil\\for time-resolved nanospintronics}} % Article title

\author{
\large
\textbf{{\L}ukasz Pawliszak${}^1$, Maria Tekielak,${}^2$ and Maciej Zgirski${}^1$}\\
%\thanks{A thank you or further information}\\[2mm] % Your name
\normalsize${}^1$Institute of Physics, Polish Academy of Sciences, al.Lotnikow 32/46, PL 02-668 Warszawa, POLAND\\
\normalsize${}^2$Faculty of Physics, University of Bia{\l}ystok, ul.Lipowa 41, PL 15-424 Bia{\l}ystok, POLAND\\ %Your institution
%\normalsize \href{mailto:john@smith.com}{john@smith.com} % Your email address
\vspace{-5mm}
}
%\vspace{-5mm}
\date{}

%----------------------------------------------------------------------------------------

\maketitle % Insert title

\thispagestyle{fancy} % All pages have headers and footers

%----------------------------------------------------------------------------------------
%	ABSTRACT
%----------------------------------------------------------------------------------------
\vspace{-0.5in}

\renewcommand{\abstractname}{}

\begin{abstract}

\noindent
Nanospintronic and related research often requires the application of fast rising magnetic field pulses in the plane of the studied planar structure. We have designed and fabricated sub-milimeter-sized coils capable of delivering pulses of the magnetic field up to $\sim500 \,\textup{Oe}$ in the plane of the sample with the rise time of order of $10 \, \textup{ns}$. The placement of the sample above the coil allows for an easy access to its surface with manipulators or light beams for, e.g., Kerr microscopy. We use the fabricated coil to drive magnetic domain walls in $1 \,  \mu\textup{m}$ wide permalloy wires and measure magnetic domain wall velocity as a function of the applied magnetic field.
\end{abstract}
%]
%----------------------------------------------------------------------------------------
%	ARTICLE CONTENTS
%----------------------------------------------------------------------------------------
\setlength{\parskip}{0cm}
\begin{multicols}{2} % Two-column layout throughout the main article text
\multicolpretolerance=100
\multicoltolerance=300
\section{Introduction}
%\vspace{-7pt}
%\lettrine[nindent=0em,lines=6]{I}

\indent A pulsed magnetic field at nanosecond timescale is used to study field-driven domain wall propagation \cite{beach_nature_materials} or to trigger magnetization reversal in magnetic nanoparticles \cite{wernsdorfer_neelbrown_model, wernsdorfer_switching}. Both phenomena are important for magnetic data storage. Pulsed magnetic fields are also useful for optical studies of magnetization or single spin relaxation in quantum semiconductor structures \cite{goryca_prb}. Furthermore, a pulsed magnetic field is of a great importance for switching "on" and "off" resonance processes in quantum systems that exhibit magnetic field-dependent separation of energy levels. For example, two qubits, when properly tuned by a magnetic field, may exchange a single photon, which leads to a creation of entangled states \cite{andreas_prl}.  The same methodology can be also employed for qubit/resonator  \cite{palacios_tunable_resonators, hofheinz_mechanical_resonator} or qubit/spin ensemble \cite{yui_prl_nvc} systems.

\indent It is relatively easy to deliver a fast rising magnetic field perpendicular to the film plane without losing the ability to observe the sample surface.  A few loop coil wound around the structure provides the required field \cite{goryca_prb}. This is a rather efficient method since a maximum of the magnetic field can easily be centered on the studied object without disturbing the access to the film plane.  However, if we want to apply the field parallel to the sample surface the situation is quite different.
Well, we can wind a coil and insert the sample with its surface parallel to the coil long axis  but at the cost of losing the direct access to the surface.  If an open access is required one has to consider another designs (Fig.\,\ref{fig:1}): (i) the structure of interest is inserted or fabricated right at the top of a stripline \cite{maekawa_prl, korn_journal_mmm}, (ii) the sample resides in the gap of the split-pair Helmholtz coil or in a horseshoe magnetic circuit or (iii) one can use the field licking outside the ordinary air-core coil or ferrite-core coil. 

In the next section we analyze the effectiveness of the above three methods. We then present our design and its optimization. This is followed by the description of minicoil assembly, testing and calibration. Next, experimental results on magnetic domain velocity are presented. Subsequently, prior to the summary, an outlook on some possible extensions of our design is presented.  
%------------------------------------------------
\section{Concepts}
\subsection{Stripline\label{subsec:stripline}}
Defining the structure under study at the top of a strip line should guarantee the best coupling  of the structure to the field since mutual distance can be kept at the level of nanometers (in case of metallic structure one needs to define a spacer between stripline and the structure). The solution seems to be ultimate for ultra fast magnetic field pulses ($\sim 1\, \textup{ns}$) \cite{korn_journal_mmm} since the inductance $L$ can be made sufficiently small and rise times for the current in the stripline scale with $L$. However, it is not always possible to fabricate such an integrated sample or it may be simply experimentally inefficient. The second limitation is related to the maximum current which may flow in a stripline due to its small thickness. One must be also conscious of the skin effects which may make stripline transport current only on the edges at sufficiently high frequencies. It follows from Ampere law that magnetic field created by the infinite plane with current density flowing uniformly in one direction is:
\begin{equation}
	\label{eq:B_max_inf_plane}
	B = \frac{\mu_0 \, j}{2}
\end{equation}
This formula gives the upper limit for magnetic field created by a stripline. The case of stripline is displayed in Fig.\,\ref{fig:1}a.
% =================================================================================
\subsection{Horseshoe magnetic circuit\label{subsec:horseshoe_magnet}}
\begin{figure}[H]
	\includegraphics[width=0.45\textwidth]{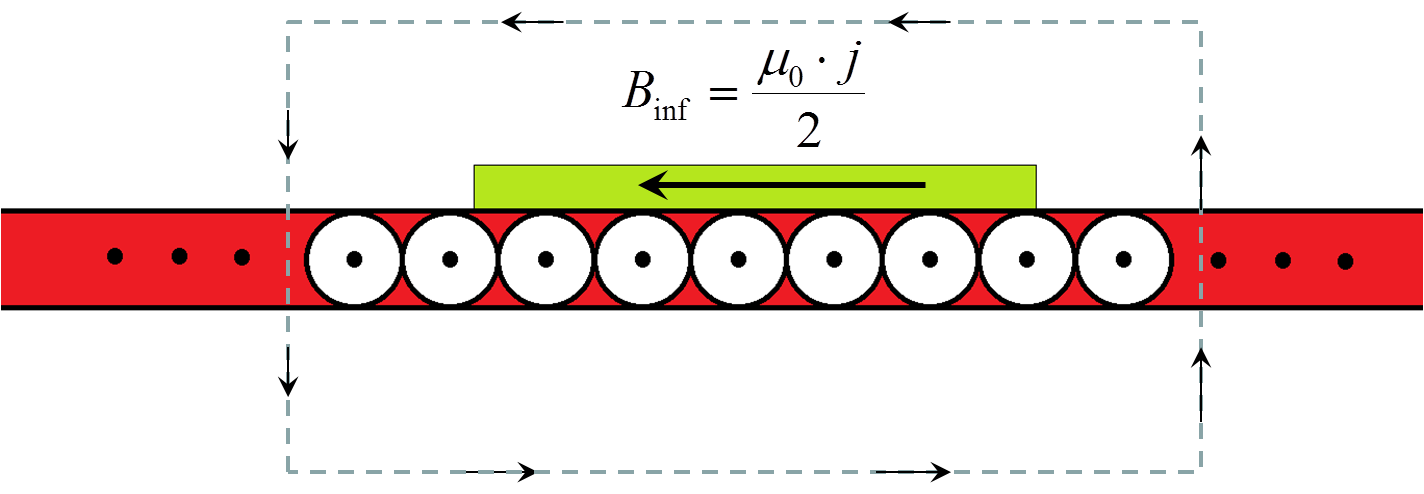}
	\put(-200,80){{\textbf{\LARGE{(a)}}}}
	\\*
	\includegraphics[width=0.45\textwidth]{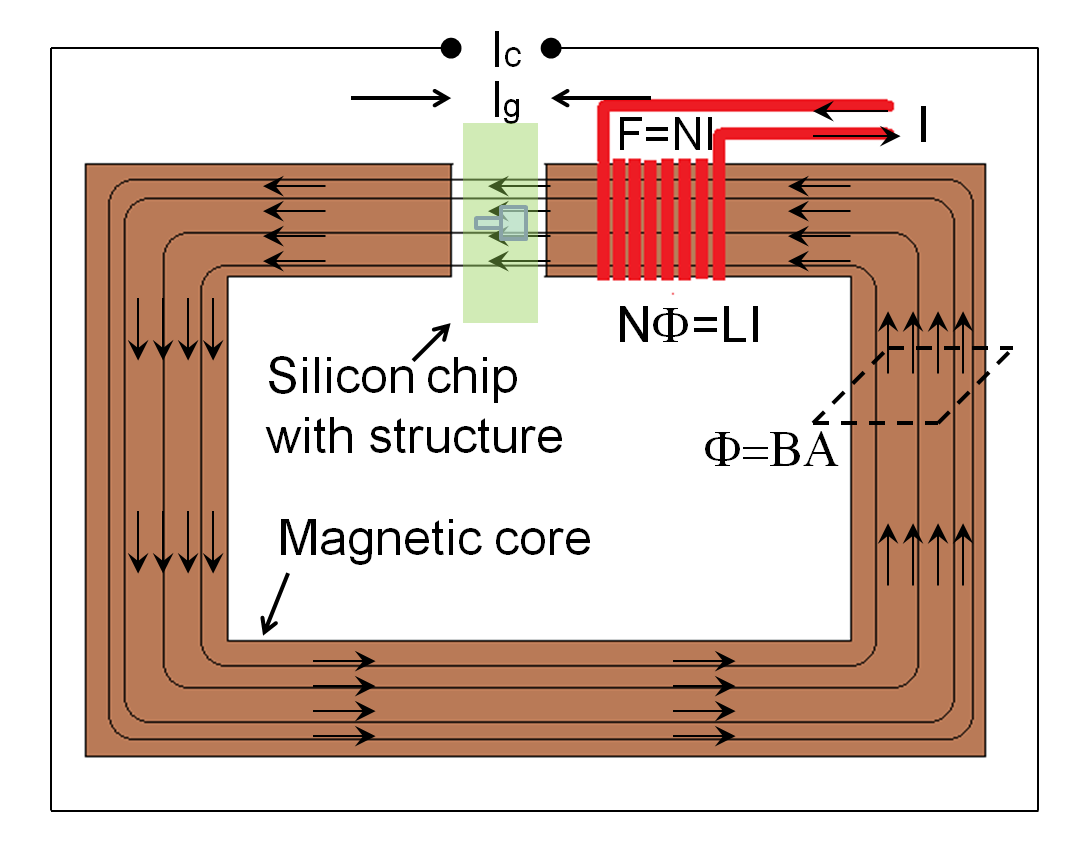}
	\put(-200,145){{\textbf{\LARGE{(b)}}}}
	\\*
	\includegraphics[width=0.45\textwidth]{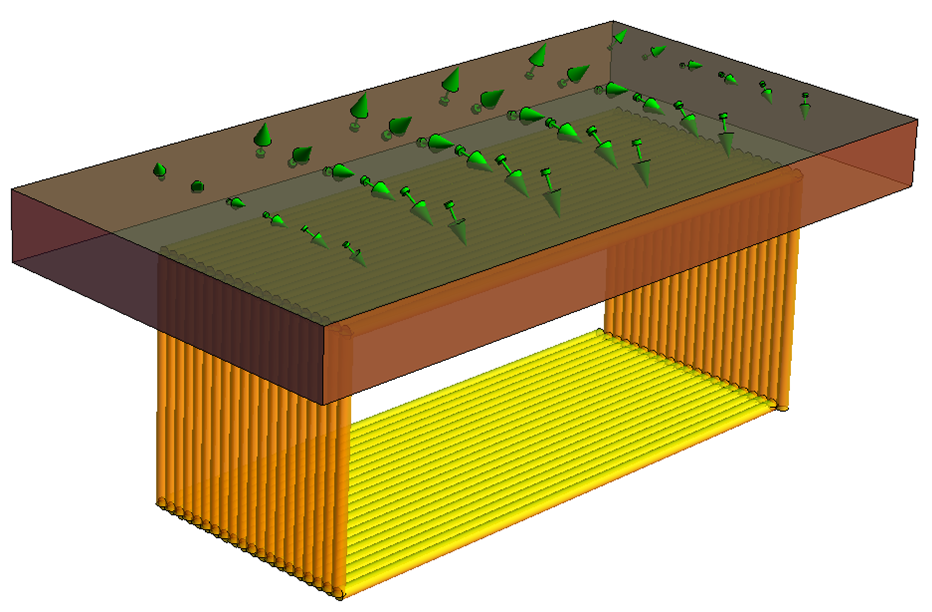}
	\put(-200,120){{\textbf{\LARGE{(c)}}}}
	\\*
	\caption{\textup{\textbf{a)}}The magnetic field generated above a stripline (red plane). It approaches the limiting value $B_{inf}$ as dimension of the stripline tends to infinity. The broken line shows the Ampere contour used to determine the magnetic field strength. Circles with dots indicate uniform current density j flowing towards observer, \textup{\textbf{b)}}The magnetic field generated in the gap of a horseshoe magnet,  \textup{\textbf{c)}}The  magnetic field generated above a coil at the surface of a silicon chip (direction of the field is depicted with vectors). The field is parallel to the surface in the middle of the assembly.}
	\label{fig:1}
\end{figure}

We take a magnetic core formed in a closed loop with a gap of length $l_g$ on the circumference to accommodate a sample (Fig.\,\ref{fig:1}b). We wind a few turns of wire $N$ somewhere around the core and let the current $I$ flow through the loops. In analyzing the described circuit it is useful to use the Hopkinson law constructed in analogy to Ohms law \cite{hopkinson_law_wiki}.
It utilizes the concept of magnetic circuit and states that magnetomotive force $\mathcal{F} = N \, I$ is a source of magnetic flux $\phi$ with reluctance $R$ being a measure of a “resistance” which magnetic circuit represents to the flux: $\mathcal{F} = \phi \, \mathcal{R}$. Reluctance is calculated in analogy to electrical resistance:
\begin{equation}
	\label{eq:horsesohe_reluctance}
	\mathcal{R} = \frac{l_c}{\mu_0\mu_r A} + \frac{l_g}{\mu_0 A}\approx\frac{l_g}{\mu_0 A}
\end{equation}
with $\mu_r\mu_0$ being a measure of magnetic flux "conductivity", $A$ – cross-section of the core, $l_c$ – length of the magnetic core along magnetic field lines inside the core, $l_g$ – gap length. The flux lines want to stay mostly in a magnetic medium since the magnetic field energy density is  $w=\frac{B^2}{2\mu_r \mu_0}$ and big $\mu_r$ guarantees lower energy. The magnetic field in the gap if $l_g < \sqrt{A}$\footnote{Condition $l_g < \sqrt{A}$ guarantees that the magnetic field inside the gap is uniform since magnetic field lines stay within the cross-section A.} reads:
\begin{equation}
	\label{eq:B_max_horseshoe}
	B = \mu_0\frac{N \, I}{l_g}
\end{equation}
And inductance of the horseshoe magnet is:
\begin{equation}
	\label{eq:L_horseshoe}
	L = \mu_0\frac{N^2A}{l_g}
\end{equation}
To get a flavor of the formulas we assume current of $1 \, \textup{A}$ and $l_g = 5 \, \textup{mm}$ (if we want to place a silicon chip with a structure in the gap $5 \, \textup{mm}$ seems to be already a small number).  As a result we obtain $2.4 \, \textup{Oe/A/turn}$. To calculate inductance we take a plausible assumption $A = {l_g}^2$  and obtain $L=6.3 \, \textup{nH}/turn^2$.

\subsection{Air-core coil}
Now we move on to the seemingly easiest design: the air-core coil (Fig.\,\ref{fig:1}c).
If such a coil is sufficiently long, formulas for the magnetic field inside the coil and the inductance are known (the Ampere law):
\begin{equation}
	\label{eq:B_max_solenoid}
	B = \mu_0\frac{N \, I}{l}
\end{equation}
\begin{equation}
	\label{eq:L_solenoid}
	L = \mu_0\frac{N^2A}{l}
\end{equation}

% =========================================================
\noindent \\
Basically these are the same formulas that we obtained for the horseshoe magnet but here length of the coil $l$ replaces the length of the gap. We can make air-core coil very small. If it has length $l=1\, \textup{mm}$ still there is no problem to place a few mm long silicon chip over it. The cross-section can also be easily reduced to a few $\, \textup{mm}^2$, say $3\, \textup{mm}^2$. With these numbers for $I=1\, \textup{A}$ we obtain a field of $12\, \textup{Oe/turn}$ inside the coil and an inductance of $3.6n\, \textup{H/turn}^2$. We notice that it is 2 times smaller inductance than in case of horseshoe magnet and at the same time $5$ times bigger field.  But it is inside long coil. Even if a field outside the coil is much smaller than inside the coil still it may be advantageous to use air-core coil instead of horseshoe magnet to create the magnetic field confined to a small volume. The next section presents numerical calculations of the field outside the coil along with influence of the coil geometry on the delivered field. The concept we will use is presented in Fig.\,\ref{fig:1}c.

% FIGURE 1  FIGURE 1  FIGURE 1  FIGURE 1  FIGURE 1  FIGURE 1  FIGURE 1  FIGURE 1  FIGURE 1  FIGURE 1  FIGURE 1  FIGURE 1  FIGURE 1

% ====================================================================
%\vspace{-8pt}
\section{Numerical design and optimization}
\begin{figure}[H]
	\includegraphics[width=0.44\textwidth]{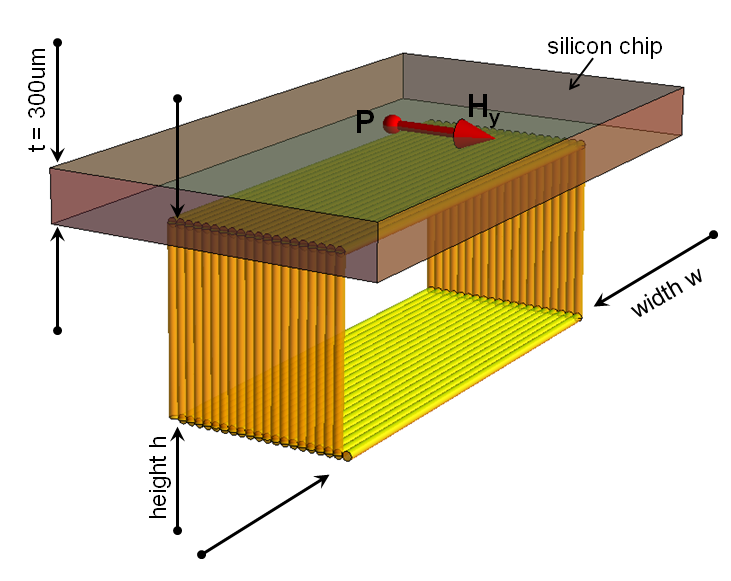}
		\put(-175,150){{\textbf{\LARGE{(a)}}}}
		\\*
	\includegraphics[width=0.44\textwidth]{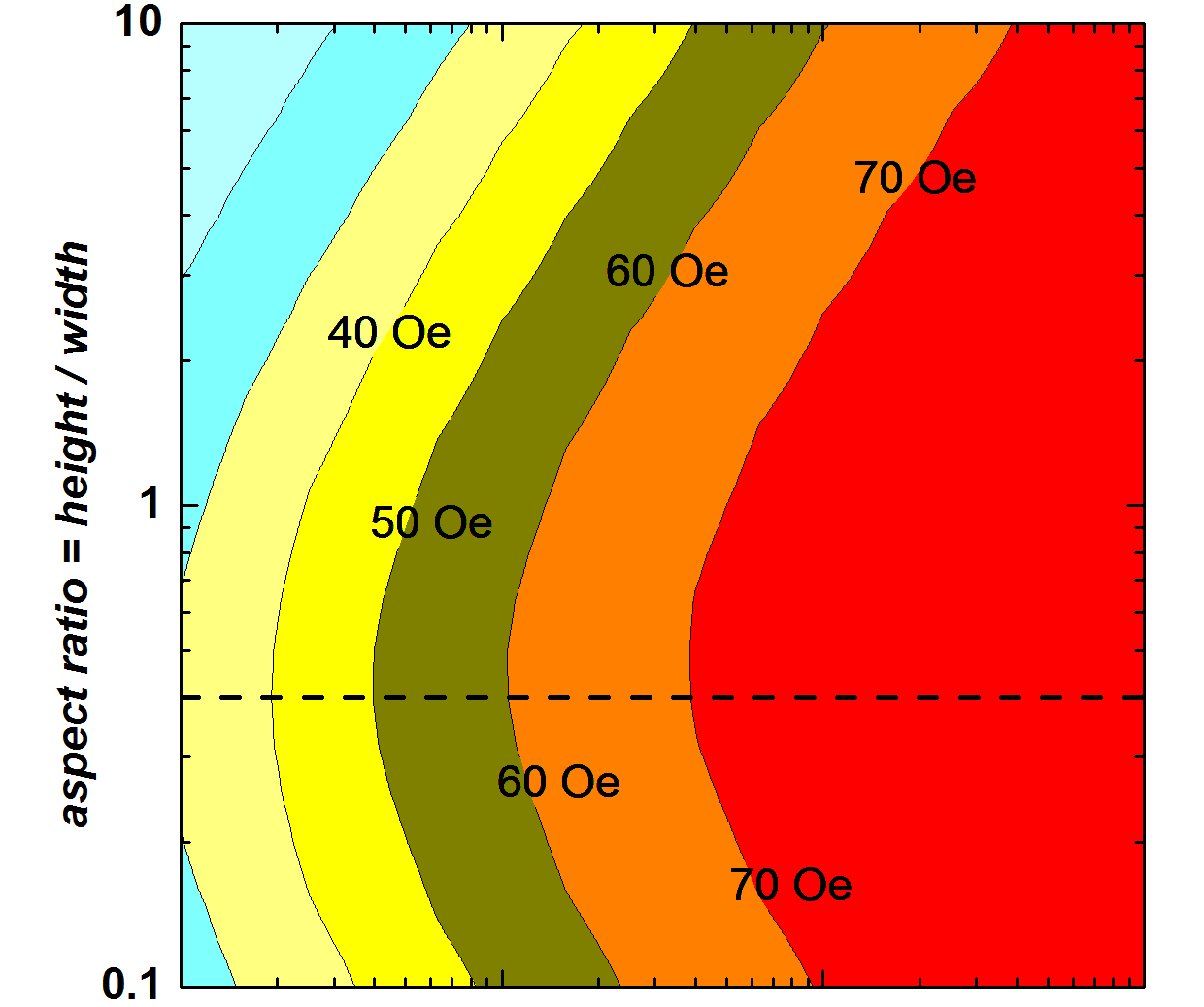}
		\put(-175,155){{\textbf{\LARGE{(b)}}}}
		\\*
	\includegraphics[width=0.44\textwidth]{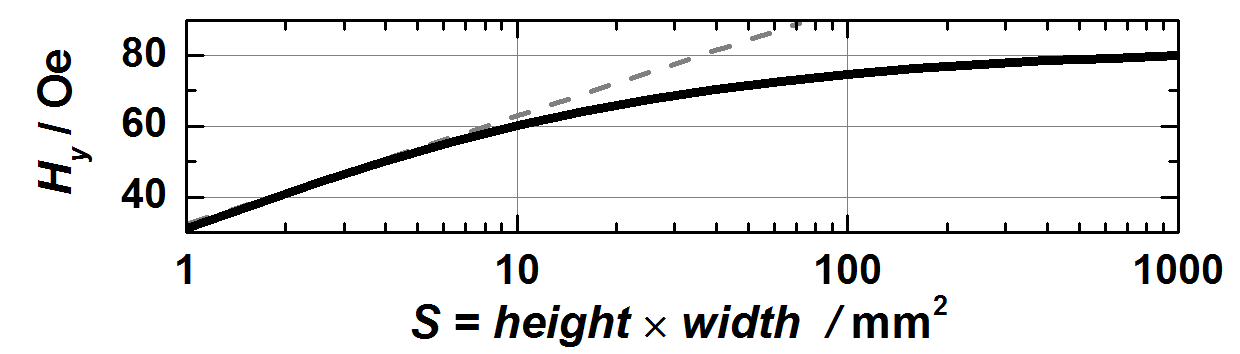}
	\put(-175,43){{\textbf{\LARGE{(c)}}}}
	\caption{Numerical calculation of the $H_y$ component of the magnetic field at point $P$ located $300\, \mu\textup{m}$ above the coil in the center of the assembly. \textup{\textbf{a)}} Geometry of the calculation, \textup{\textbf{b)}} $H_y$  generated for various geometries and sizes of the coil driven with current of $1\, \textup{A}$. Length of the coil $1\, \textup{mm}$, \textup{\textbf{c)}} Section of the field map from b) along dashed line.}
	\label{fig:2}
\end{figure}

It may seem that a field outside coil is too small to be considered (in fact for long solenoid it is the field that we neglect when using Ampere law to calculate inductance and the field inside coil). Nevertheless it is possible to optimize geometrical dimensions of the coil in order to get the maximum magnetic field outside the coil in direction parallel to coil main axis. Such an optimization must be done numerically but before showing it up we go on for a while with  qualitative considerations. \\
\indent
	We have already presented the case of infinite current plane in section~\ref{subsec:stripline} together with the formula for the magnetic field above the plane. It reads: $B = \frac{\mu_0 j}{2}$. Now we consider a coil with rectangular cross-section like the one presented in Fig.\,\ref{fig:2}a. It consists of  $4$ “walls” with current. If we place sample very close to one of the ”walls” and the wall is sufficiently large we start to approach the case of infinite plane. If we assume that other $3$ “walls” are far away from the sample we can neglect the field they produce.  Suppose we wound the coil with $50\mu\textup{m}$ Copper wire and drive it with $1\, \textup{A}$ current.  The linear current density is $j = 1\, \textup{A/}50\mu\textup{m} = 20 \, \textup{A/mm}$. It follows that the limiting value for the field produced by the coil is $H_{inf}\approx 125.6\, \textup{Oe}$. We expect to approach this number as we increase dimensions of the coil.\\
\indent
We analyze influence of the geometry and size of the coil on the field produced outside the coil in the center $300\mu\textup{m}$ above it (Fig.\,\ref{fig:2}a) for the coil with fixed length of $1\, \textup{mm}$ wound with 1 layer of $50\mu\textup{m}$ wire (20 turns), but with different widths and heights.

To find the field we build a coil out of many segments (i.e. 2000 segments per turn) with $\vec{r}_i$ and $\vec{r}_{i+1}$ specifying ends of a segment and use the Biot-Savart law in the numerical form:
\begin{equation}
	\label{eq:biot_savart_law}
	\begin{array}{cc}
		d\vec{B} = \frac{\mu_0}{4\pi}\frac{I}{r^3}\left(\vec{r}_i - \vec{r}_{i+1}\right)\times\vec{r},
		
	\end{array}
\end{equation}
with $\vec{r} = \vec{r}_p - \frac{\vec{r}_i + \vec{r}_{i+1}}{2}$ defining position of a point where we want to find the field  $\vec{r}_p$ with respect to the middle of a segment. The result of the numerical integration of the Biot-Savart law over entire coil (i.e. over i) is displayed in Fig.\,\ref{fig:2}b.
It shows that increasing size of the coil (its cross-section) has only logarithmic effect on the produced field, for bigger sizes the effect is even smaller. It also shows that flat coils (with height to width ratio of $\sim\textup{0.4}$) represent the best trade-off between strength of the field and coil dimensions (which we want to minimize to reduce inductance). When cross-section of the coil is fixed such a geometry is the best compromise between fields produced by upper and lower “walls” of the coil which are opposite in direction outside the coil: for flatter coils lower "parasitic" wall starts to have more influence on the field, for coils with bigger aspect ratio the area of the upper ”wall” becomes smaller producing less field.  The coil obviously cannot approach the mentioned limit for the infinite plane since its length is fixed and equal to $1\, \textup{mm}$. However increasing the length of the coil (see Supplementary Material, Fig.\,\ref{fig:S3}) reveals the asymptotic behavior.

%FIGURE2 FIGURE2FIGURE 2 FIGURE2FIGURE2FIGURE2FIGURE2FIGURE2FIGURE2FIGURE2FIGURE2FIGURE2FIGURE2FIGURE2FIGURE2

\begin{figure}[H]
	\includegraphics[width=0.45\textwidth]{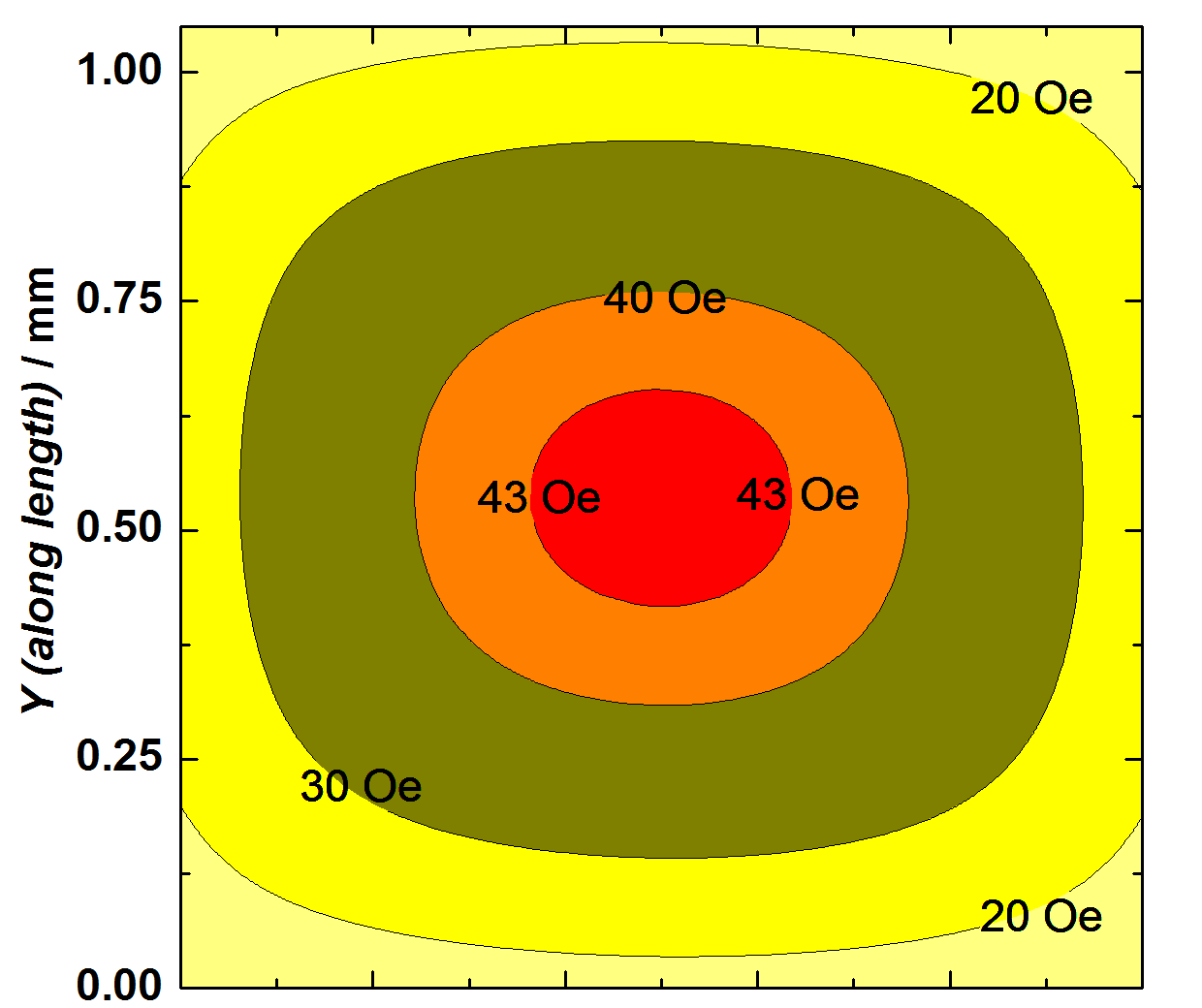}
		\put(-110,160){{\textbf{\LARGE{(a)}}}}
		\\*
	\includegraphics[width=0.45\textwidth]{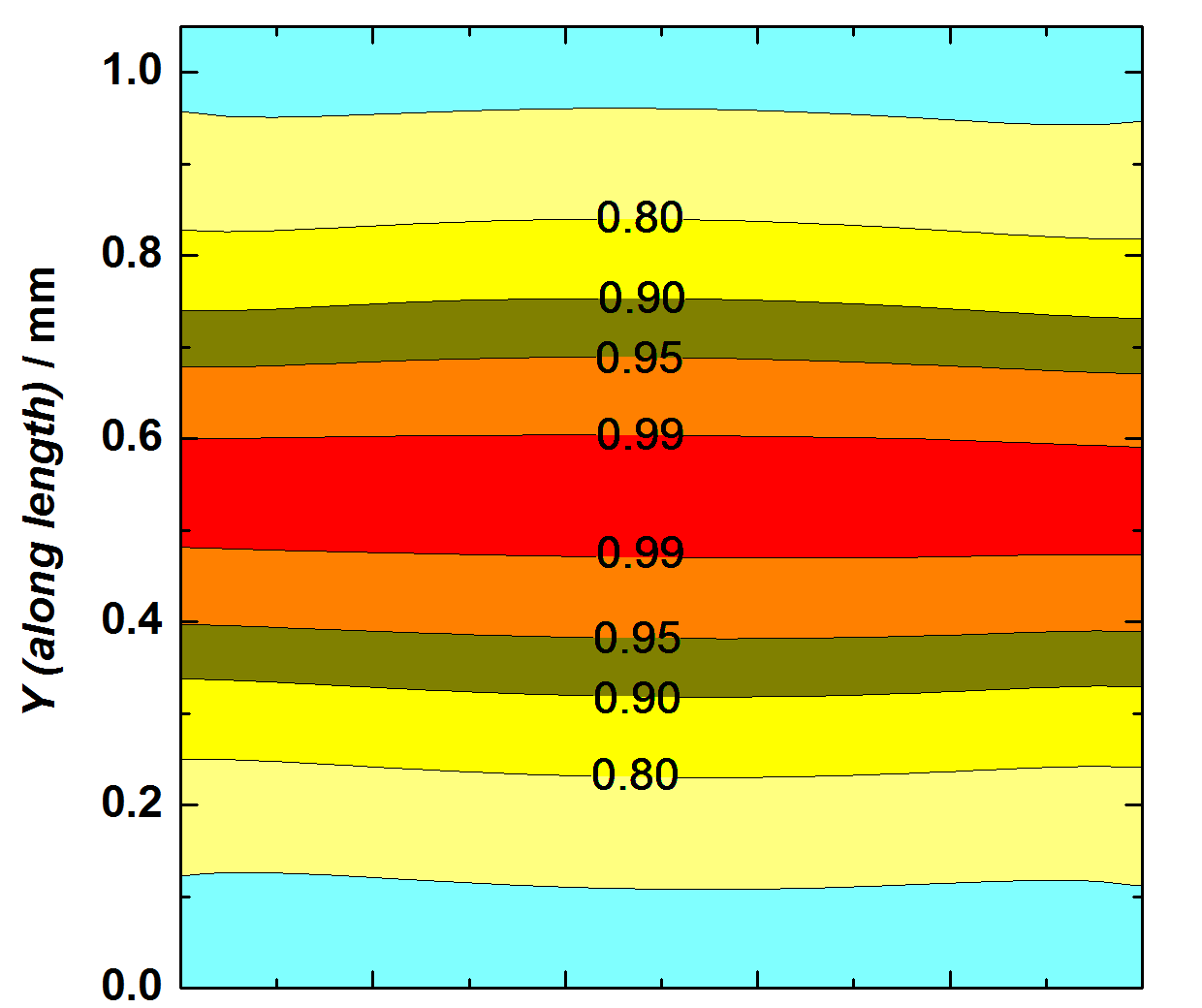}
	\put(-110,160){{\textbf{\LARGE{(b)}}}}
	\\*
	\includegraphics[width=0.45\textwidth]{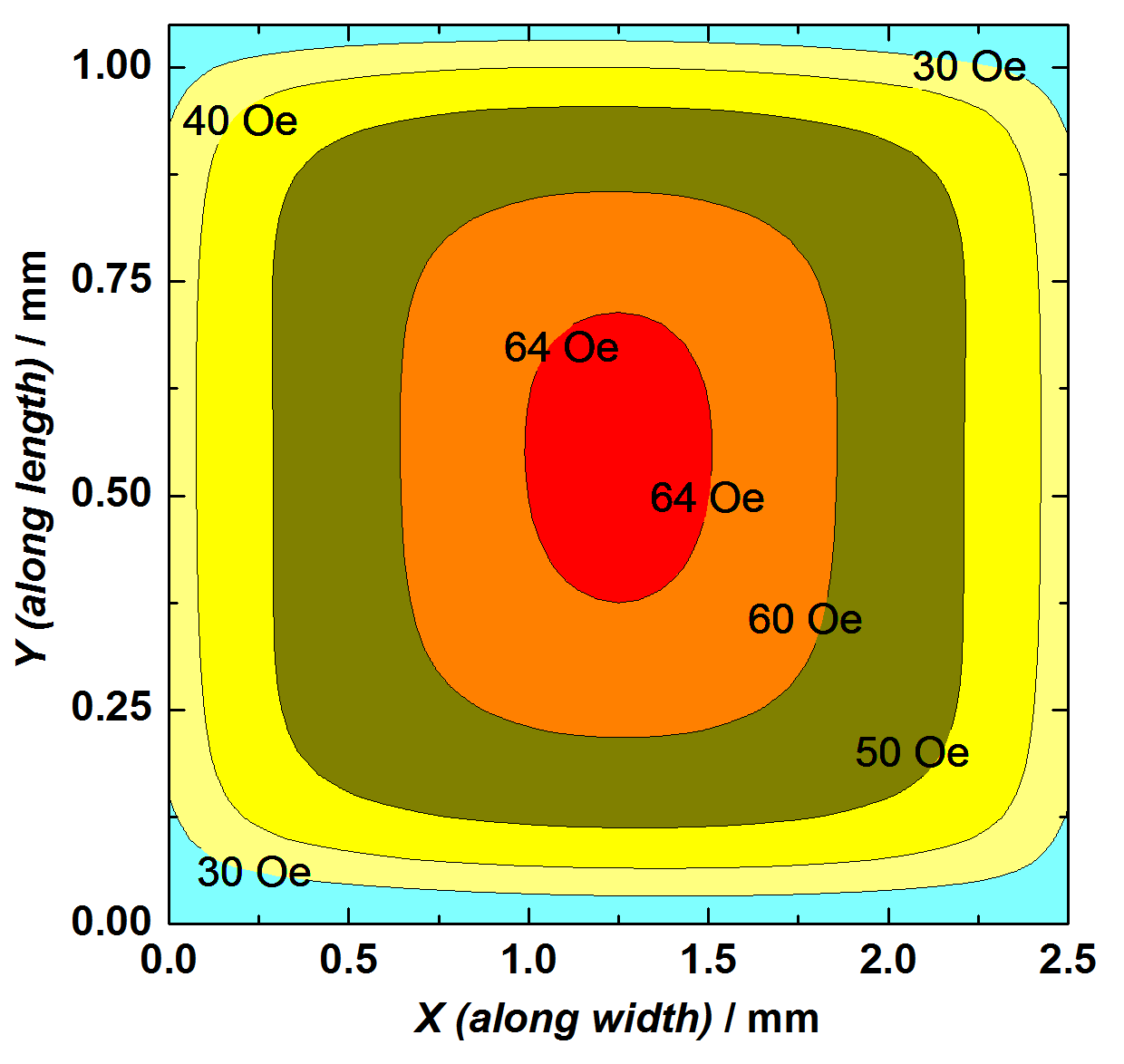}
		\put(-110,180){{\textbf{\LARGE{(c)}}}}

	\caption{\textup{\textbf{a)}} Variation of $H_y$ component of the field in the plane placed  $300 \, \mu\textup{m}$ above the coil, \textup{\textbf{b)}} The amount of $H_y$ component of the field in the total field produced in the plane placed  $300 \, \mu\textup{m}$ above the coil, \textup{\textbf{c)}} Variation of $H_y$ component of the field in the plane placed $100 \, \mu\textup{m}$ above the coil.}
	\label{fig:3}
\end{figure}

Next we focus on the particular coil with length of $l=1.05\, \textup{mm}$, (21 turns), width $w = 2.5\, \textup{mm}$ and height $h = 1\, \textup{mm}$. Its aspect ratio is $h/w = \textup{0.4}$. Being based on calculation from Fig.\,\ref{fig:2}b we expect it produces a field in excess of $40\, \textup{Oe}$ for $1\, \textup{A}$ current $300\mu\textup{m}$ above the coil in the center (cf. Fig.\,\ref{fig:2}a).
We analyze how uniform is the field in the plane placed $300 \, \mu\textup{m}$ above the coil (it is the plane where the sample is to be placed) by calculating the value of $H_y$ component of the field. The relevant map is presented in Fig.\,\ref{fig:3}a.

We observe that there is a sizable spot above the center of the coil where the field remains relatively constant. In Fig.\,\ref{fig:3}b we compare the component $H_y$ to the total field generated in the plane. Over sizable strip the $H_y$ dominates. The relative homogeneity and strength of the field is even larger in the plane $100 \, \mu\textup{m}$ above the coil (Fig.\,\ref{fig:3}c).
One should  use thin substrates (e.g. $100 \,\mu\textup{m}$ thick Si wafer) if uniformity of the field is an issue. We notice that our air-core coil generates higher field than horseshoe magnet (Fig.\,\ref{fig:1}b) described in section \ref{subsec:horseshoe_magnet} if inductances of the two are made the same by choosing proper number of turns in the horseshoe magnet.
We also observe that obtaining the same field for a $500 \,\mu\textup{m}$ wide stripline (wide enough to create the homogeneous field over a few $\sim100 \times 100\, {\mu\textup{m}}^2$) would require application of a $10$ times larger biasing current than in the case of air-core coil (a coil allows to circulate “the same current” over many loops).

%FIGURE3

% ========================================================================================
\section{Minicoil assembly, testing and calibration}
\begin{figure}[H]
	\includegraphics[width=0.49\textwidth]{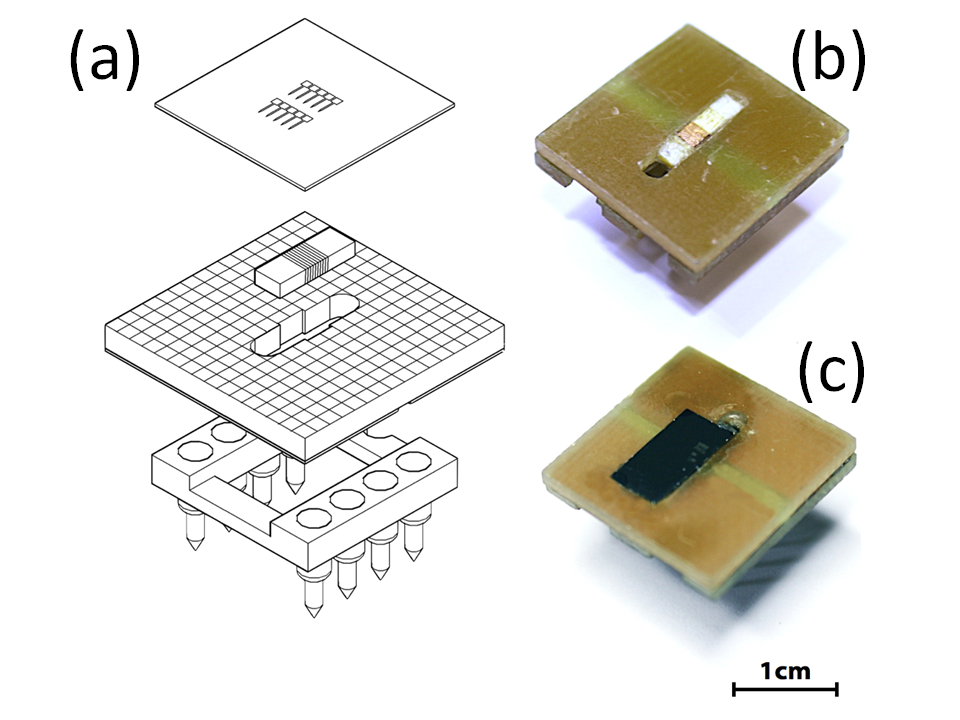}
	\caption{Magnetic coil assembly. \textup{\textbf{a)}} Exploded view of the coil assembly with a magnetic coil and a PCB with a cut to accomodate the coil and grating on the surface of the PCB allowing for a precise placement of a sample. The coil is connected to 2 electrical pins of a standard 8-pin base (bottom feature) and delivers the field to a sample placed over it (e.g. defined on silicon chip – seen at the top), \textup{\textbf{b)}} A magnetic coil wound on a piece of a wooden stick of $1\times2.5\, \textup{mm}^2$ cross-section mounted in a PCB, \textup{\textbf{c)}} Silicon chip with some structures placed exactly over the coil with help of the grating.
}
	\label{fig:4}
\end{figure}

\begin{figure*}
	\includegraphics[width=0.5\textwidth]{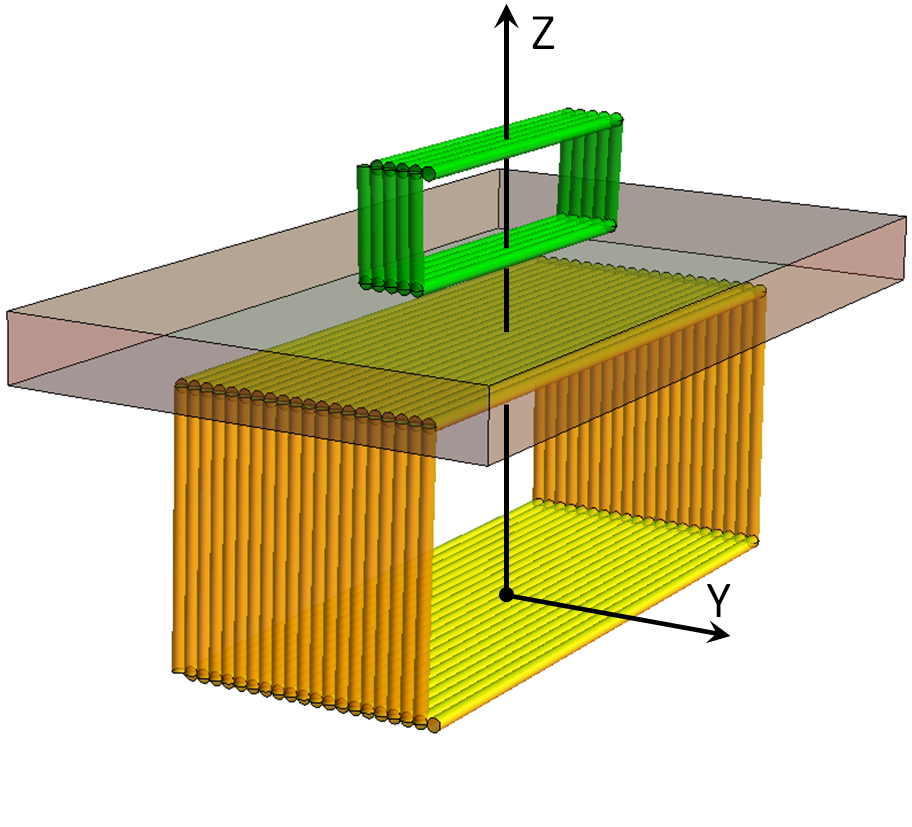}
	\put(-220,180){{\textbf{\LARGE{a)}}}}
	\includegraphics[width=0.5\textwidth]{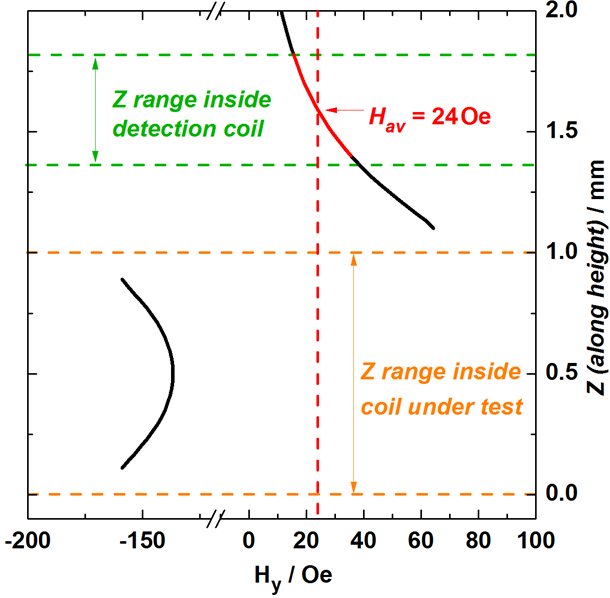}
	\put(-220,180){{\textbf{\LARGE{b)}}}}
	\caption{\textup{\textbf{a)}} The geometry used to calibrate the coil. The detection coil (smaller one) is placed over the coil-under-test with $300 \, \mu\textup{m}$ thick silicon chip separating two coils, \textup{\textbf{b)}} Numerical calculation of the $H_y$ component of the magnetic field generated by the coil-under-test ($\sim1\times2.5 mm^2$ cross-section) biased with current of 1A  along z-axis crossing the coil in the middle. The average magnetic field through detection coil is $24\, \textup{Oe/A}$.}
\label{fig:5}
\end{figure*}

\begin{figure*}
\centering
	\includegraphics[width=0.5\textwidth]{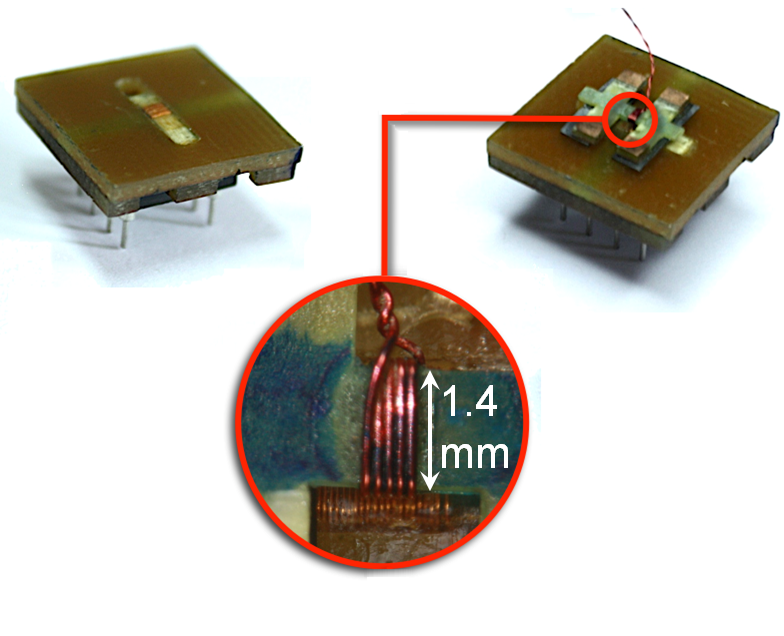}
	\put(-220,180){{\textbf{\LARGE{a)}}}}
	\put(-110,180){{\textbf{\LARGE{b)}}}}
	\includegraphics[width=0.5\textwidth]{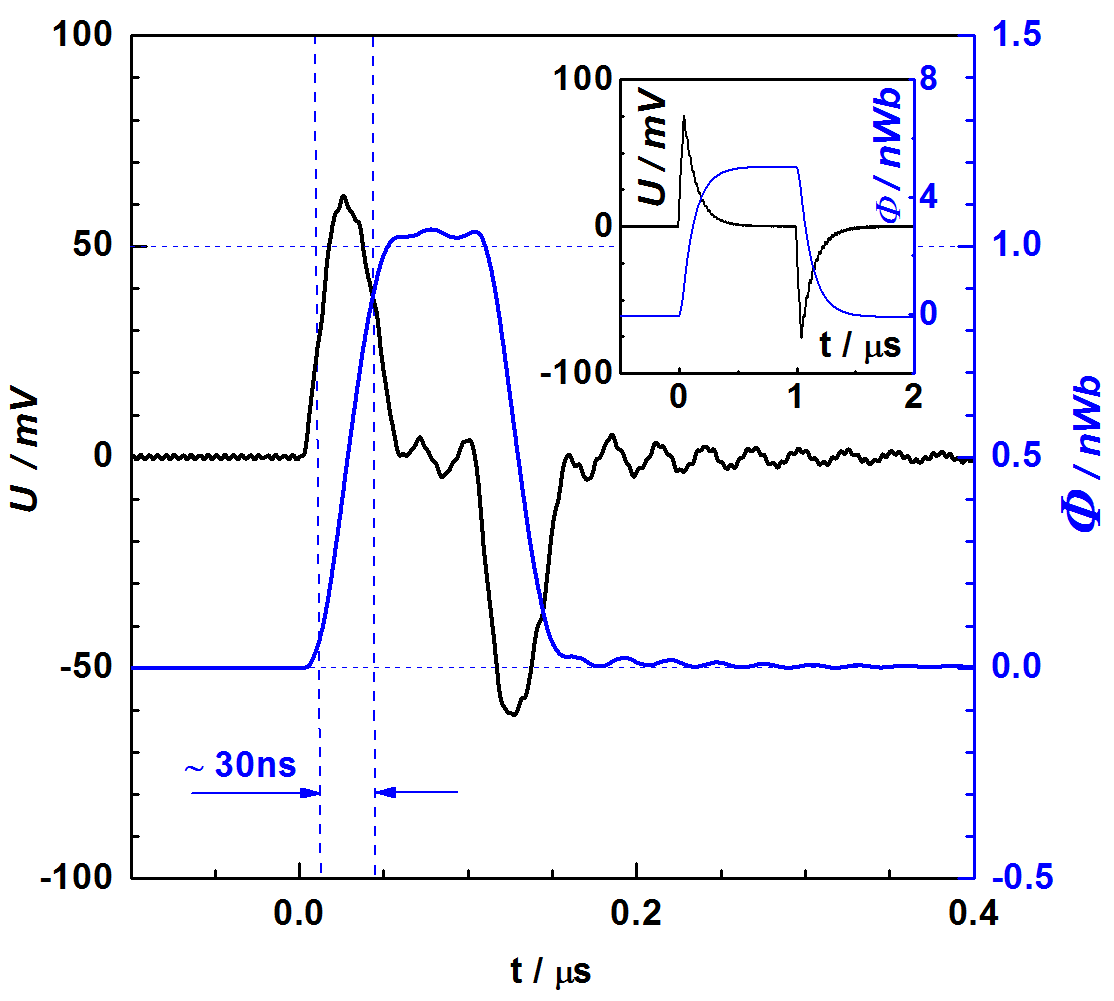}
	\put(-200,180){{\textbf{\LARGE{c)}}}}
	\caption{Determination of the field delivered by the coil. \textup{\textbf{a)}} Coil-under-test, \textup{\textbf{b)}} Coil-under-test with detection coil placed over it. The close-up shows the \emph{5}-loop detection coil (wire diameter = $100\, \mu\textup{m}$, cross-section $0.4\times1.4\, \textup{mm}^2$) sitting above the coil-under-test,  \textup{\textbf{c)}} The voltage signal generated in the detection coil from Fig.\,\ref{fig:6}b in response to current pulse of magnitude $188\, \textup{mA}$ going through the coil-under-test (left axis) and corresponding magnetic flux in the detection coil obtained on integrating the voltage signal (right axis). The inductance of the coil under study is $L=0.8\, \mu\textup{H}$. Coil was biased through $R=50\Omega$ resistor ($\tau=\sfrac{L}{R} = 16\, \textup{ns}$). Inset shows signal generated by a similar coil, but with ferrite core. 5 fold increase in the magnetic field strength is accompanied by significant degradation of the field rising time (see section 6.2 for more details).}
\label{fig:6}
\end{figure*}

Following the way of reasoning presented so far we constructed a minicoil assembly (Fig.\,\ref{fig:4}).
We wound coils on matchsticks of  rectangular cross-sections of $\sim2-3 mm^2$ with a few tens of turns. Inductances of obtained coils were characterized by measuring the impedance modulus vs. frequency and fitting parameters $L$ and $R$ of the series circuit of which coil was a part (Supplementary Material, Fig.\,\ref{fig:S2}). Values of inductances were at the level of $\sim1\, \mu\textup{H}$. The value of the magnetic field above the coils-under-study was measured by AC-coupling to even smaller detection coil placed $300\, \mu\textup{m}$ above. The idea and its experimental realization are presented in Fig.\,\ref{fig:5} and Fig.\,\ref{fig:6}.
A varying current flowing through coil-under-study is the source of the alternating magnetic field and induces the EMF in the detection coil. On integrating the EMF we obtain the magnetic flux through detection coil. In Fig.\,\ref{fig:6}c we present the magnetic flux pattern in the detection coil  in response to $\sim100\, \textup{ns}$ long square current pulse  with rise time of $\sim30\, \textup{ns}$ and amplitude of $188\, \textup{mA}$ triggered from Agilent 33250A into the coil-under-test ($\sim1\times2.5 mm^2$ core, 21 turns with Cu $50\, \mu\textup{m}$ wire).  The flux captured by the coil inevitably exhibits vertical variation with the strongest field at $\sim300 \, \mu\textup{m}$ above the coil (in place where the actual sample is to be placed if standard $300\, \mu\textup{m}$ thick Si substrate is used). Nevertheless it is possible to compare it with numerical calculation.
In Fig.\,\ref{fig:5}b we display numerical calculation off the $y$ component of the field $H_y$ generated by coil-under-test along Z-axis crossing the coil in the middle (cf. Fig.\,\ref{fig:5}a). In the range from $z = 1.35\,\div2.0\,\textup{mm}$, dependently on the exact placement of the detection coil, we obtain the average magnetic field $H_y^{av} = (24 \pm 5)$\,Oe/A inside the coil. The total flux threading the detection coil reads: $\phi= N\, Area\, B_y^{av} $ (for the values of parameters see \ref{fig:6}b) and should be equal to $(6.7 \pm 2.1)$\,nWb for the $1\, \textup{A}$ current in the coil-under test (in addition to the uncertainty of the average magnetic field within the detection coil we determined uncertainty of the Area threaded by the flux to be $10\, \textup{\%}$, see Supplementary Material). From experiment we got 5.4\,nWb/A (Fig.\,\ref{fig:6}c). Considering the difficulty to place the detection coil at exact position above the coil-under-test, its possible slight spatial shift and uncertainty in determination of cross-sections of both coils (see Supplementary Material) we obtained satisfactory agreement.

% ====================================================================================
\section{Dynamic control of magnetic domain walls in permalloy nanowires}
\begin{figure}[H]
	\centering
	\includegraphics[width=0.42\textwidth]{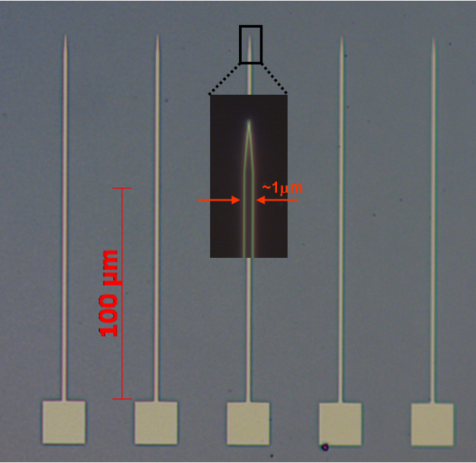}
	%\rule{\linewidth}{3cm}
	\caption{Optical image of 5 nominally identical $1\mu\textup{m}$ wide $20\, \textup{nm}$ thick Permalloy nanowires.}
	\label{fig:7}
\end{figure}

We have used constructed coil to dynamically control magnetic domain walls movement in Permalloy nanowires. Such wires are expected to be inherent part of future spintronic devices, serving as a medium for encoding information or even performing logic operations \cite{allwood_science}.
$20\, \textup{nm}$ thick nanowires were prepared by conventional e-beam lithography on $100\, \mu\textup{m}$ thick Si substrate followed by thermal evaporation of Permalloy (\ce{Ni80Fe20}) at base pressure of $2\times10^{-7}\, \textup{mBar}$ (Fig.\,\ref{fig:7}).\\

%FIGURE 7 FIGURE 7 FIGURE 7 FIGURE 7 FIGURE 7 FIGURE 7 FIGURE 7 FIGURE 7 FIGURE 7 FIGURE 7 FIGURE 7 FIGURE 7 FIGURE 7 FIGURE 7

\begin{figure}[H]
	\centering
	\includegraphics[width=0.47\textwidth]{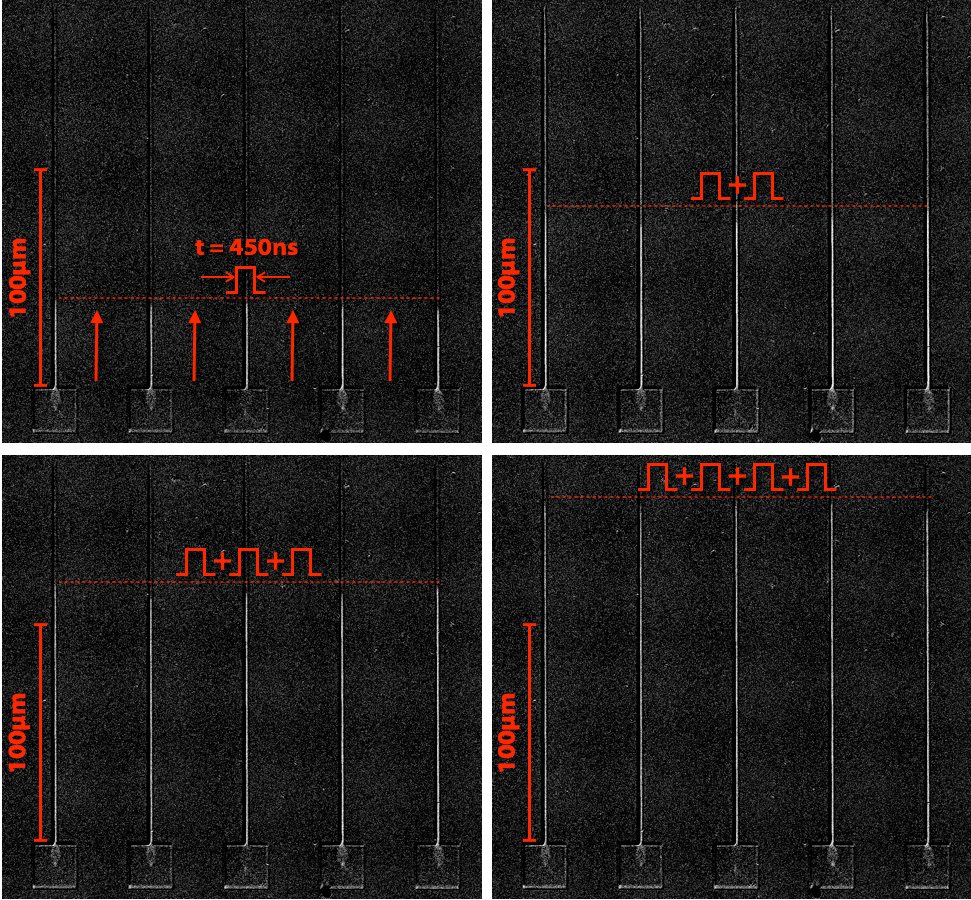}
	%\rule{\linewidth}{3cm}
	\caption{Domain-walls observed with Kerr microscopy in 5 nominally identical $1\mu\textup{m}$ wide $20\, \textup{nm}$ thick Permalloy nanowires after application of 1, 2, 3 and 4 identical $450\, \textup{ns}$ long magnetic field pulses.}
	\label{fig:8}
\end{figure}

The wires were connected at one side to $20\times20\mu\textup{m}^2$ permalloy pads, on the other - they were terminated with tips. Such a geometry allows for easy nucleation of magnetization reversal in pads (at low field values) and propagation of the domain wall from the pads \cite{ono_apl}. Tips prevent magnetization reversal nucleation \cite{schrefl_magn_mater}.\\
\indent
The substrate was placed on a minicoil with nanowires placed parallel to the magnetic field generated by the coil. Permalloy exhibits shape anisotropy \cite{maekawa_prl, erskine_prb}, with easy axis parallel to the length of a wire. The anisotropy becomes larger with narrowing of the wire. By applying short pulses of the magnetic field we could follow the position of the magnetic domain wall with longitudinal magneto-optical Kerr effect (L-MOKE).\\
\indent
We calibrated minicoil against commercial magnet by analyzing shift of 2 hysteresis loops collected on permalloy pads: one with constant current of $186\, \textup{mA}$ in the minicoil, another with current of $-186\, \textup{mA}$ (the current was delivered from Agilent 33250A and measured with Fluke multimeter connected in series to the coil). From the calibration we obtained the field of $(40\pm10)\, \textup{Oe / A}$. For further studies we used home-built current generator capable of delivering of $14\, \textup{A}$ to the coil in short pulses. The $50\, \mu\textup{m}$ Copper wire, from which the coil was wound, could sustain constant current up to $0.5\, \textup{A}$ (burning limit), but in short pulses ($\sim3 \,\mu\textup{s}$) it proved to withstand even $14\, \textup{A}$.\\
\indent
Below we briefly outline a procedure we have used for controlling domain wall position. First we saturate a nanowire in the constant field of $1.5\, \textup{kOe}$ of a big commercial magnet applied along the length of the wire. Then we reverse the direction of the field and set its value to zero. Afterwards the magnetic field pulse from minicoil is applied. In the Fig.\,\ref{fig:8} we present Kerr microscopy “stroboscopic” pictures with clear  positions of magnetic walls after applying 1, 2, 3, 4 successive identical $450\, \textup{ns}$ long magnetic field pulses of $\sim23\, \textup{Oe}$ amplitude.

%Figure 8 Figure 8 Figure 8 Figure 8 Figure 8 Figure 8 Figure 8 Figure 8 Figure 8 Figure 8 Figure 8 Figure 8 Figure 8 Figure 8 Figure 8
\noindent
The  reproducibility of the data is proven by the same advancement of the domains in 5 neighboring nanowires.

\begin{figure}[H]
\centering
	\includegraphics[width=0.49\textwidth]{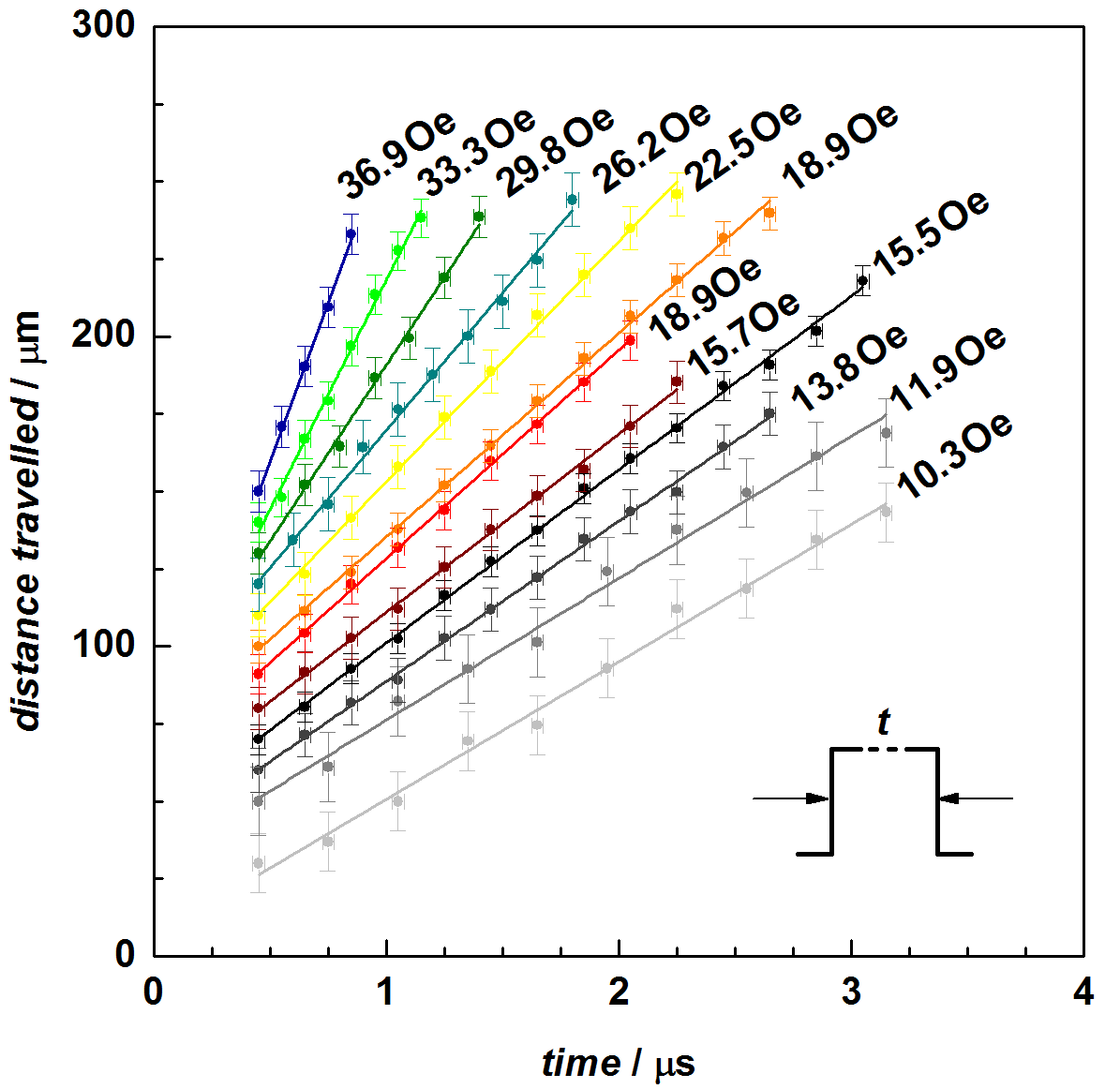}
	\caption{Distance traveled by domain wall as a function of magnetic pulse duration for various fields. Each point is averaged over 4 trials for 5 wires presented in Fig.\,\ref{fig:7}. The error bars correspond to variation of domain wall position from trial to trial. The solid lines are the least-square fits, whose slopes give the domain wall velocity displayed in Fig.\,\ref{fig:10}. The data were offset for clarity.}
\label{fig:9}
\end{figure}

\par
\begin{figure}[H]
\centering
	\includegraphics[width=0.49\textwidth]{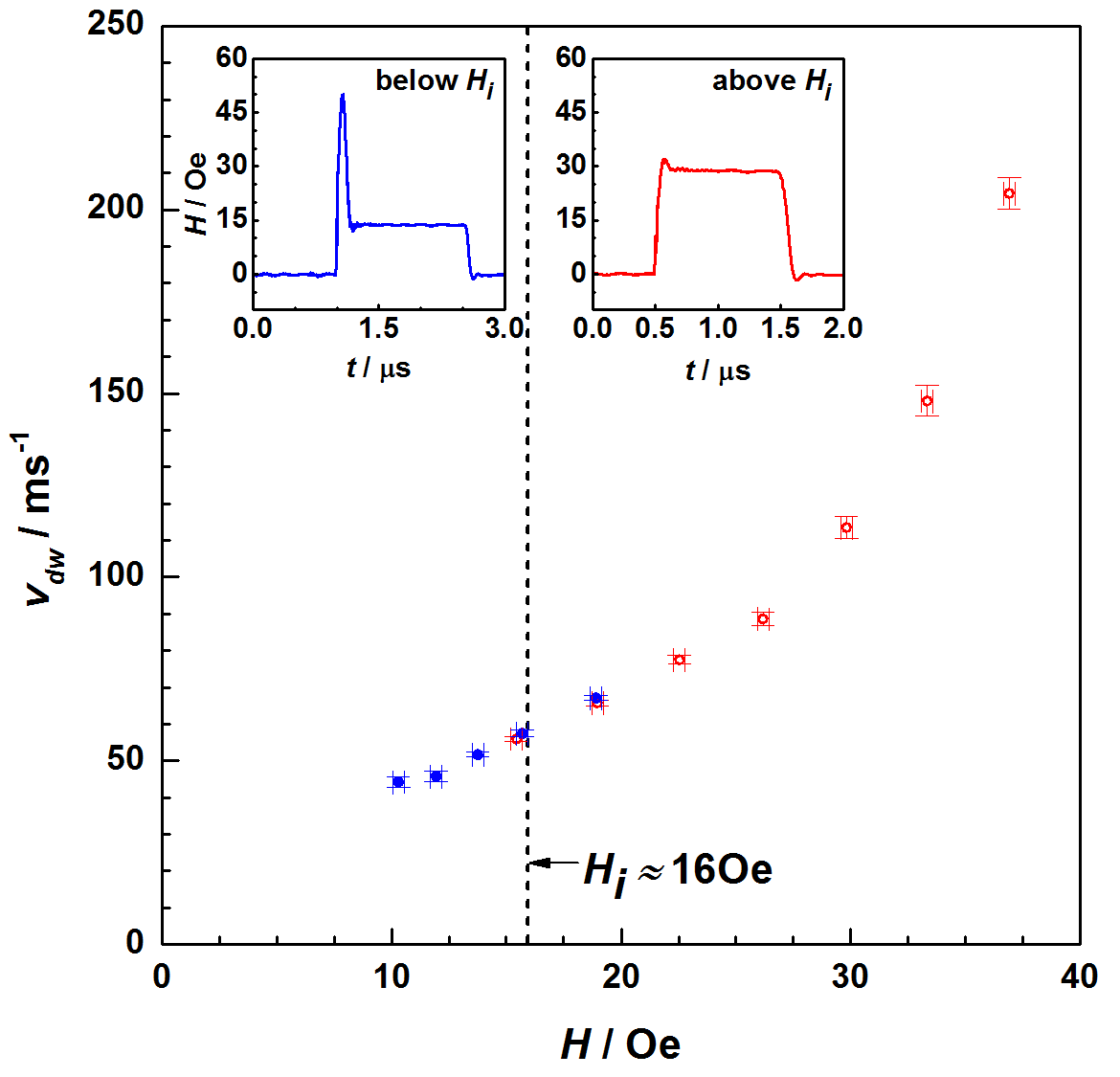}
	\caption{Domain wall velocity versus field amplitude. The domain wall injection field $H_i$ is indicated with dashed vertical line. Open symbols were obtained for a square-wave drive field (right inset). Filled symbols show velocities obtained with so called injection pulse (with constant duration of $\sim100\, \textup{ns}$) proceeding the actual field plateau (left inset). The overlap of 2 points obtained with first pulse with 2 velocities obtained with second pulse shows consistency of the data. The relative systematic error in the field calibration is estimated to be $25\, \textup{\%}$.}
\label{fig:10}
\end{figure}

To measure wall velocity at a given magnetic field we progressively increased the duration of the magnetic field pulse collecting 4 pictures for each pulse,  providing enough statistics to determine uncertainty of the data (Fig.\,\ref{fig:9}). Each image was processed with ImageJ software \cite{imagej} to extract the distance domain wall traveled.
After each pulse the wire was “reset” with $1.5\, \textup{kOe}$, as already described. For fields above $16\, \textup{Oe}$ we used rectangular pulse to drive the domains. Below this value, so called injection field \cite{beach_nature_materials, erskine_prb}, we found it difficult to provide with reproducible launching of the wall from the pad. To circumvent the problem we employed short prepulse \cite{beach_nature_materials, erskine_prb} injecting the domain wall into wire.
Once the wall was injected it could move along the wire in fields below the injection threshold. We could follow domain movements reliably down to $10\, \textup{Oe}$. Below this value the movement was not reproducible with big variation of a traveled distance from trial to trial.  Velocities determined at different fields (slopes of the straight lines from Fig.\,\ref{fig:9}) are displayed in Fig.\,\ref{fig:10}.

%Figure 9 Figure 9 Figure 9  Figure 9 Figure 9 Figure 9 Figure 9 Figure 9 Figure 9 Figure 9 Figure 9 Figure 9 Figure 9 Figure 9 Figure 9 Figure 9 Figure 9 Figure 9

%\vspace{12pt}
Our results are complementary to those reported in literature \cite{beach_nature_materials, ono_science} where detailed studies of the wall movement in $600\, \textup{nm}$ and $500\, \textup{nm}$ Permalloy nanowires were presented.  In future we are planning to use minicoils to test spin-motive force arising from the magnetization reversal \cite{maekawa_prl, erskine_prb}.

%Figure 10 Figure 10 Figure 10 Figure 10 Figure 10 Figure 10 Figure 10 Figure 10 Figure 10 Figure 10 Figure 10 Figure 10 Figure 10 Figure 10 Figure 10 Figure 10
% =======================================================================================
\newpage
\section{Alternatives}
\subsection{Vector field minicoil}
\begin{figure}[H]
\includegraphics[width=0.23\textwidth]{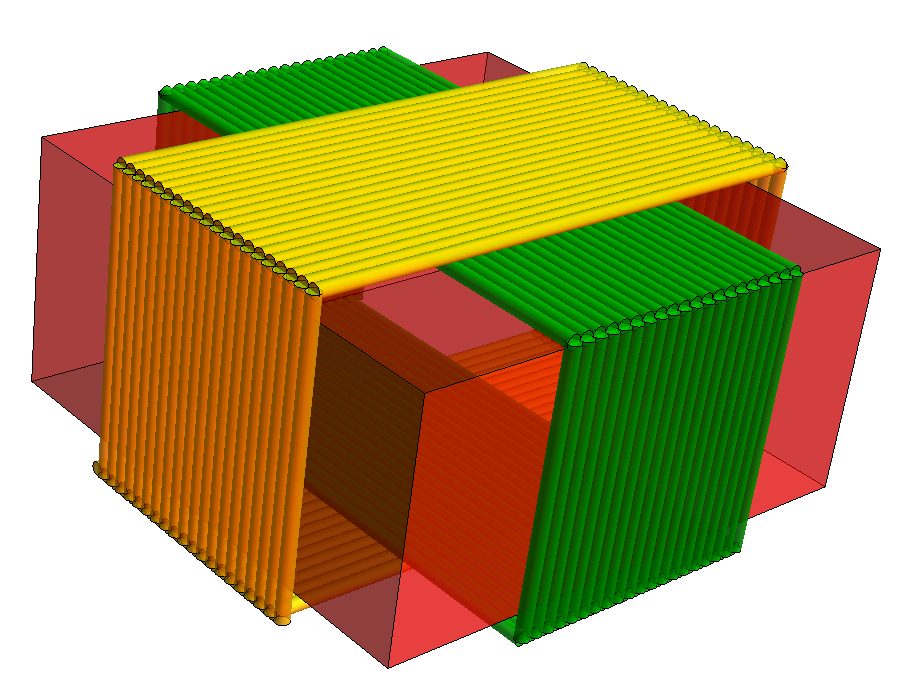}
\put(-100,90){{\textbf{\LARGE{(a)}}}}
\includegraphics[width=0.23\textwidth]{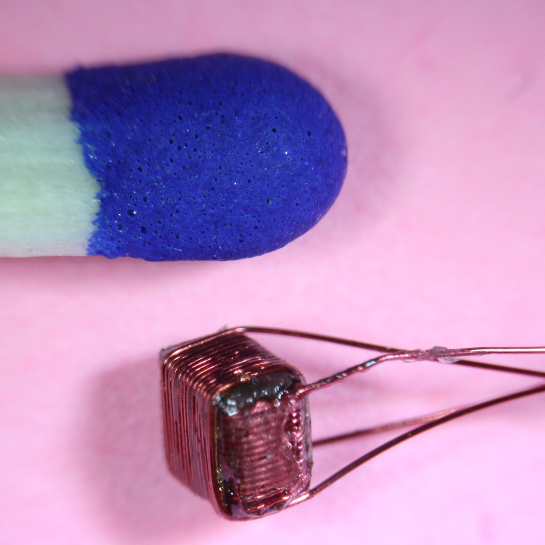}
\put(-30,90){{\textbf{\LARGE{(b)}}}}
\caption{ \textup{\textbf{a)}} 2D vector magnet design,  \textup{\textbf{b)}} its practical realization photographed together with a head of a matchstick.}
\label{fig:11}
\end{figure}

By winding two coils perpendicular to each other one obtains a 2D vector magnet (Fig.\,\ref{fig:11}).

%FIGURE 11 FIGURE 11 FIGURE 11 FIGURE 11 FIGURE 11 FIGURE 11 FIGURE 11FIGURE 11FIGURE 11FIGURE 11FIGURE 11FIGURE 11FIGURE 11

\subsection{Coil with ferrite}
\begin{figure}[H]
\centering
	\includegraphics[width=0.49\textwidth]{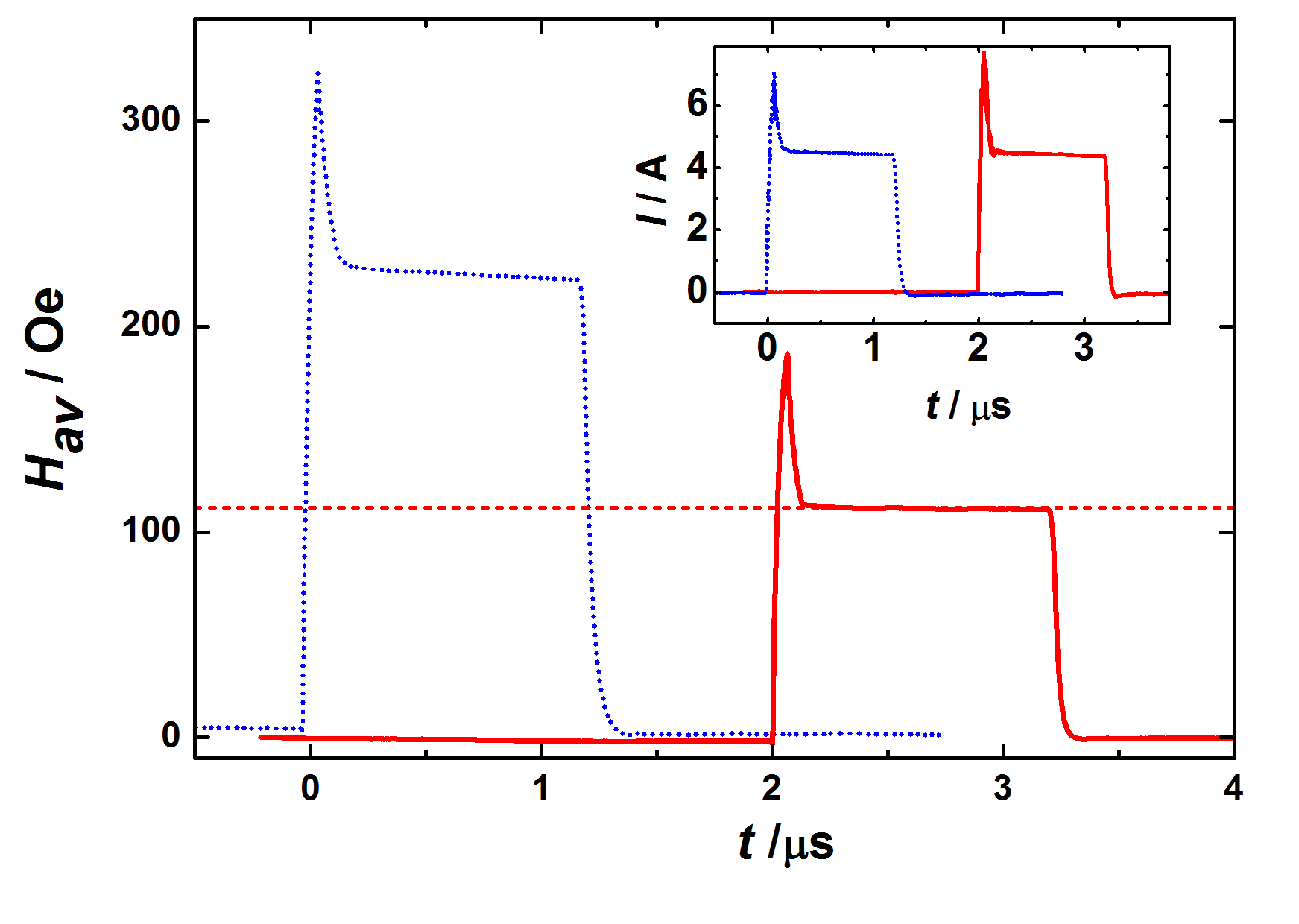}
	%\rule{\linewidth}{3cm}
	\caption{The average field generated by two coils-under-test in detection coils - one with ferrite (left, dotted line), another with teflon (right, solid line) core. Both coils were biased with the same current pulses (inset: dotted - with ferrite, solid - with teflon). By proper engineering we could double the field without compromising the speed.}
\label{fig:12}

\end{figure}

We can significantly increase the strength of the magnetic field in the discussed geometry by using a magnetic core, i.e., ferrite material \cite{ferrite_catalog}.  The field strength and inductance of the coil will be both magnified by factor $\mu_{eff}$ – effective magnetic permeability:
\begin{equation}
	\label{eq:B_max_coil_w_ferrite}
	B = \mu_0\mu_{eff}\frac{ N \, I}{l}
\end{equation}
\begin{equation}
	\label{eq:L_coil_w_ferrite}
	L = \mu_0\mu_{eff}\frac{N^2A}{l}
\end{equation}
where $N$ is number of turns, $l$ - length of the coil covered with turns and $A$ -- cross-section of the coil.
If we wind the coil on a magnetic rod with length $l$ comparable to its diameter obtained magnetic circuit is open and effective magnetic permeability of the coil becomes much smaller than the relative permeability of the rod.  For example for commercial ferrite materials $\mu_{eff}$  in such a case is only $3.5\div4$ \cite{ferrite_catalog}, but it already gives a significant enhancement of the field. If we keep the same length of the coil but reduce the number of turns by factor of  $\sqrt{\mu_{eff}}$ the inductance of the coil will remain the same compared to the air core coil but the strength of the field will increase by factor of  $\sqrt{\mu_{eff}}$.\\
\indent
In the inset of Fig.\,\ref{fig:6}c we present the magnetic field generated in the detection coil by a non-optimized coil-under-test with ferrite core.  For the same current as for the air-core coil and the same detection coil magnetic flux is 5 times bigger but  inductance also has increased from $1\, \mu\textup{H}$ to $5\, \mu\textup{H}$. It results in much worse rising time. However by using idea presented above one can take advantage of a ferrite material without compromising the speed of the coil.
We have produced two cores: one made of Teflon, another made of ferrite 4B1 with $\mu_{eff}\sim4$ \cite{ferrite_catalog}. Both were shaped into $\sim2.5\times1.5\, \textup{mm}^2$ cross-section cuboids. Teflon core was wound with 20 turns of $50\, \mu\textup{m}$ wide \ce{Cu} wire. Ferrite core with 10 turns of $100\, \mu\textup{m}$ \ce{Cu} wire. Thus the lengths of the cores covered with wires were the same. We tested obtained coils by attaching detection coils at the top of coil assemblies as shown in Fig.\,\ref{fig:5}a. Both detection coils were prepared on $1.35\times0.53\, \textup{mm}^2$ cross-section cores by winding $5$ turns of $100\, \mu\textup{m}$ wide \ce{Cu} wire and glued to the $250\, \mu\textup{m}$ thick substrate sitting on the top of coil assemblies (Fig.\,\ref{fig:6}). We measured the fields generated in two cases for the same biasing current-pulse and found roughly double enhancement of the field strength for ferrite-based coil (Fig.\,\ref{fig:12}). At the same time no degradation of the rising time was observed, proving the idea.
%\begin{figure}[H]
%\includegraphics[width=.5\textwidth]{Fig11}
%\caption{This is a caption}
%\label{fig:11}
%\end{figure}

\subsection{Inverted Helmholtz coil}
By winding a coil with a gap in the middle (or winding two coils close to each other) one can increase surface above the coil where the magnetic field is uniform. It is similar to a split pair Helmholtz coil but this time we use the field outside the coil.\\
\par

\section{Summary}
In conclusion, we have designed, built, and validated the universal magnetic field mini-source. It can be used in any area of science, in which fast rising magnetic field pulses are required to test tiny objects. The coil delivers a field in the film plane without compromising the access to the surface and allows for an easy replacement of samples. We have tested the design by determining the velocity of magnetic domains in permalloy wires as a function of the external magnetic field.

\section*{Acknowledgements}
The authors are grateful to Foundation for Polish Science for supporting the work through the HOMING PLUS program. We thank T.\,Dietl, A.\,Maziewski for helpful discussions and A.\,Stupakiewicz and P.\,Czerwi\'nski for support in measurements.
%----------------------------------------------------------------------------------------
%	REFERENCE LIST
%----------------------------------------------------------------------------------------

%----------------------------------------------------------------------------------------
\appendix
\addtocounter{figure}{-12}
\makeatletter

\renewcommand{\thesubsection}{S\@arabic\c@subsection}
\renewcommand{\thefigure}{S\@arabic\c@figure}
\renewcommand{\thetable}{S\@arabic\c@table}
\makeatother

\newpage

\section*{Supplementary material}
\subsection{Detection coil calibration}
\begin{figure}[H]
	\includegraphics[width=0.5\textwidth]{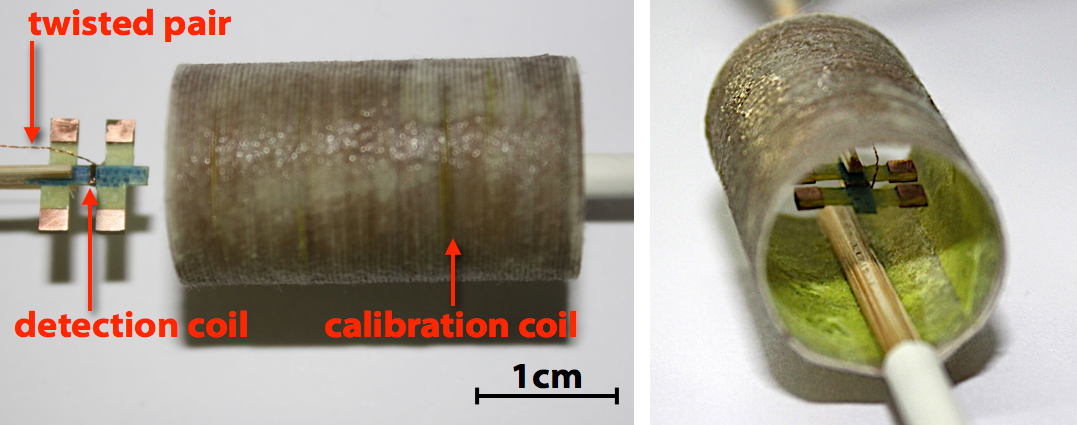}
	\put(-120,80){{\textbf{\LARGE{(a)}}}}
	\put(-30,80){{\textbf{\LARGE{(b)}}}}
	\caption{Detection coil calibration. \textup{\textbf{a)}} Big coil - the source of magnetic field  and
small coil wound on a piece of a PCB with rectangular cross-section of $1.4\times0.4\, \textup{mm}^2$
for detection of the varying magnetic field by means of the Faraday law,
 \textup{\textbf{b)}} Detection coil in a uniform field of the big coil.}
	\label{fig:S1}
\end{figure}

Our detection coil has an area of the order of $0.5\, \textup{mm}^2$. Its linear dimensions are comparable with diameter of the wire used to wind the coil.  It follows that exact cross-section of the coil is a problematic number to determine. We wound big coil approaching the limit of long solenoid and put the detection coil inside it (Fig.\,\ref{fig:S1}). The field generated inside the big coil is well described with the Ampere law. With sinusoidal signal of known amplitude applied to the big coil we could calculate the effective area of the detection coil by measuring the amplitude of the resulting magnetic flux. The cross-section determined in this way remains within $10\, \textup{\%}$ in agreement with the cross-section of the core measured with a caliper.
%FIGURE S1 FIGURE S1FIGURE S1FIGURE S1FIGURE S1FIGURE S1FIGURE S1FIGURE S1FIGURE S1FIGURE S1FIGURE S1FIGURE S1FIGURE S1FIGURE S1

\subsection{Determination of inductance of a minicoil}
\begin{figure}[H]
\begin{center}
	%\begin{circuitikz}[scale=.8]
		%\draw (0,0)
		%to[short, -*] (0,2)
		%to[R=$R_m$] (2,2)
		%to[L=$L$] (4,2)
		%to[R=$r$,-*] (6,2)
		%to[R, l=50<\ohm>] (8,2)
		%to[sV, l=$Agilent$] (8,0)
		%to[short, -*](6,0)
		%to[short](0,0);
		%\draw(0,2)
		%to[short](0,4)
		%to[voltmeter, l=$V_m \text{ = } I_m\cdot R_m$](2,4)
		%to[short, -*](2,2);
		%\draw[<->] (6,0.25)  to[bend right] node[anchor=west]{$U$} (6,1.75);
		%%\draw[<->] (6,0.25)  -- node[anchor=east]{$U_m \text{=} I_m\cdot  Z$} (6,1.75);
	%\end{circuitikz}
	%\includegraphics[width=.5\textwidth]{FigS2a}
	\includegraphics[width=0.5\textwidth]{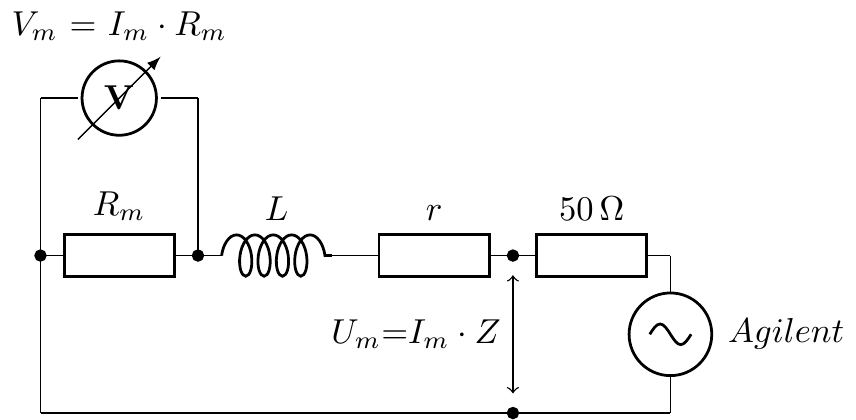} %obwd w pdf
	\put(-100,80){\textbf{\LARGE{(a)}}}
		%\put(0,0){{a)}}
\\*
	\includegraphics[width=0.5\textwidth]{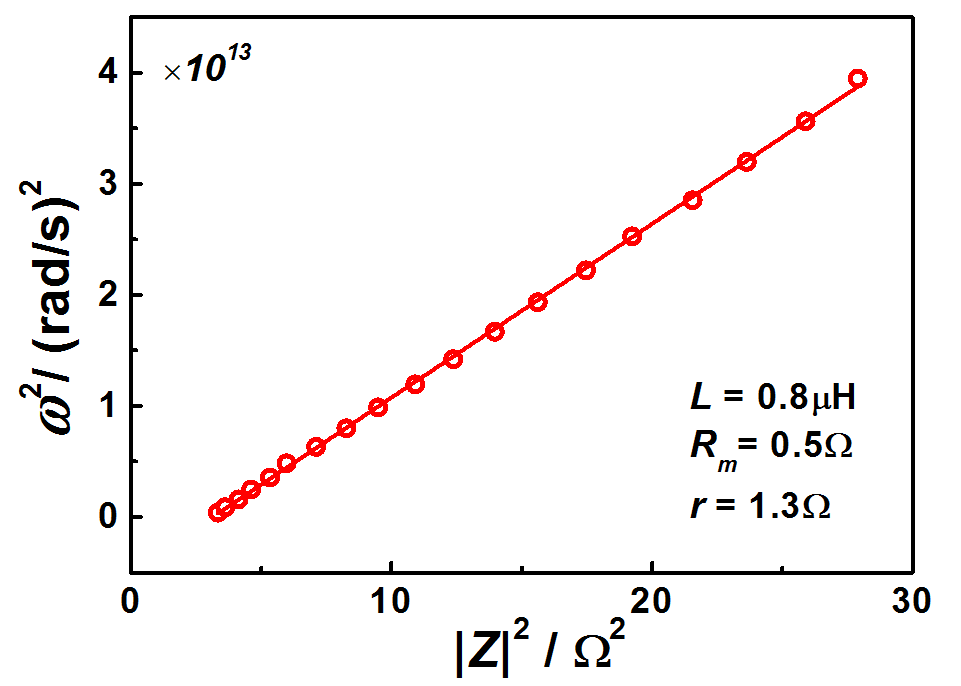}
		%\put(0,0){{b)}}
			\put(-100,140){{\textbf{\LARGE{(b)}}}}
    	\caption{\textup{\textbf{a)}} Circuit used for determination of the coil inductance. $V_m$ is voltage measured with oscilloscope on the resistor of the known value $R_m = 0.5\,\Omega$, $U_m$ is the voltage measured across the circuit, $50\,\Omega$ is the internal resistance of Agilent 33250A generator, \textup{\textbf{b)}} Experimental data (dots) displayed along with least square fit.}
	\label{fig:S2}
\end{center}
\end{figure}

Inductances of minicoils have been characterized by measuring the impedance modulus vs. frequency for the series circuit presented in Fig.\,\ref{fig:S2}a. Total impedance of the circuit reads: $Z=R+i\omega L$. Experimentally it is measured as a ratio $U_m/I_m$. On comparing the two and taking the squared modulus we arrive at the following formula:
\begin{equation}
	\label{eq:omega_2}
	\omega^2=\frac{1}{L^2}\left|\frac{U_m}{I_m}\right|^2 - \frac{(R_m + r)^2}{L^2}
\end{equation}
By fitting a straight line we determine inductance $L$ of the coil and series resistance $R=R_m+r$. A representative data and fit are displayed in Fig.\,\ref{fig:S2}b. The outlined procedure gives the value of $L$ in full agreement with direct measurement by means of the \emph{RLC bridge Hameg - HM8118}, but in a wider frequency range.
\subsection{Big coil approaching the field limit of an infinite plane}
\begin{figure}[H]
	\includegraphics[width=0.5\textwidth]{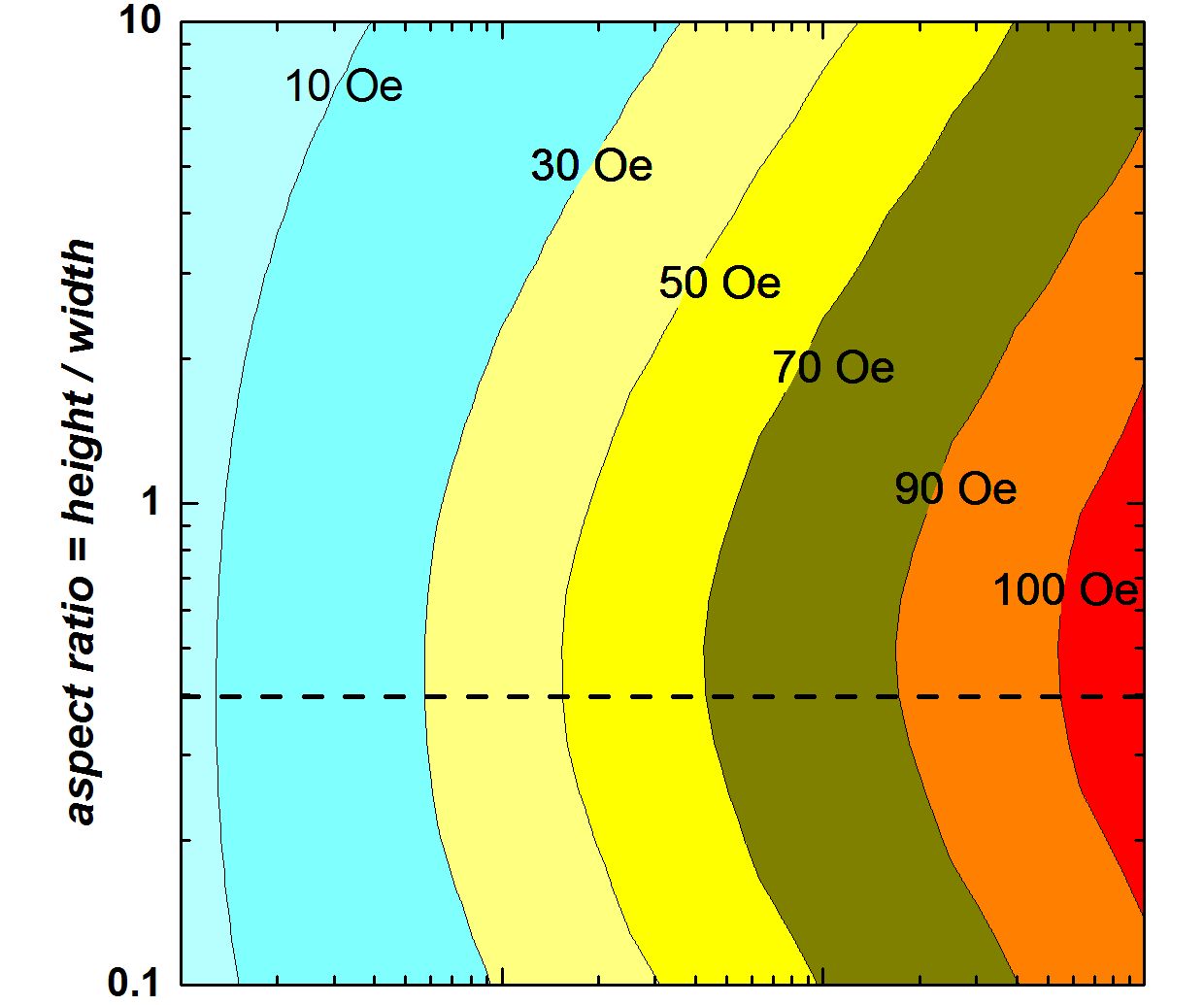}
	\put(-200,160){{\textbf{\LARGE{(a)}}}}
	\\*
	\includegraphics[width=0.5\textwidth]{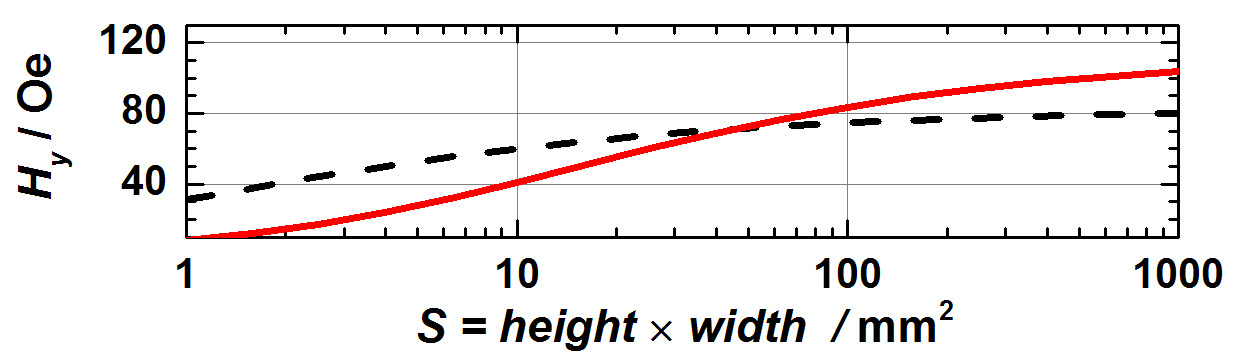}
	\put(-200,50){{\textbf{\LARGE{(b)}}}}
	\caption{\textup{\textbf{a)}} $H_y$  generated for various geometries and sizes of the coil driven with current of $1\, \textup{A}$. Length of the coil $4\, \textup{mm}$, \textup{\textbf{b)}} Solid line represents section of the field map from a) along the dashed line. For comparison the dashed line shows profile for smaller coil ($1\, \textup{mm}$ long) also presented in Fig.\,\ref{fig:2}c. Note that shorter coil produces larger field for smaller cross-sections and longer coil wins for larger dimensions.}
	\label{fig:S3}
\end{figure}

It has been argued in the main text that linear current density of $20\, \textup{A}/\mu\textup{m}$ in an infinite plane should give rise to the field of $125.4 \, \textup{Oe}$ above the plane. In Fig.\,\ref{fig:S3} we present numerical calculation of the field generated $300 \, \mu\textup{m}$ above the coil (the geometry is presented in Fig.\,\ref{fig:2}a) as a function of its height and width but with fixed length of $4\, \textup{mm}$ corresponding to 80 turns wound with $50\, \mu\textup{m}$ wide wire. It is clear that increasing size of the coil we start approaching the limiting value for the infinite plane.

\end{multicols}

\end{document}